\documentclass[pra,aps,superscriptaddress,groupedaddress]{revtex4}
\usepackage{xspace,amsmath,amsfonts,amsthm,amssymb,amsbsy}
\usepackage{graphicx}
\usepackage{amssymb}
\usepackage{color}
\usepackage{psfrag}
\usepackage{ifsym}
\usepackage{float}
\usepackage{epstopdf}
\usepackage[usenames,dvipsnames]{xcolor}
\usepackage[colorlinks=true,citecolor=Cerulean,linkcolor=RubineRed,urlcolor=Cerulean]{hyperref}
\usepackage{bbold}
\usepackage{cancel}

\usepackage{rotating} 

\usepackage{slashbox} 

\definecolor{pink}{rgb}{1,0.078,0.57}

\definecolor{green}{rgb}{0,0.7,0.9}

\newcommand{\Tr}{\mathrm{Tr}}

\newcommand{\mf}{\mathbf}

\newcommand{\beq}{\begin{equation}}
	\newcommand{\eeq}{\end{equation}}
\newcommand{\beqa}{\begin{eqnarray}}
	\newcommand{\eeqa}{\end{eqnarray}}

\newcommand{\eqa}[1] {\begin{eqnarray} \centering #1 \end{eqnarray}}

\newcommand{\av}[1]{\left \langle #1 \right \rangle}

\newcommand{\iu}{i}

\newcommand{\la}{\langle}
\newcommand{\ra}{\rangle}

\begin{document}

		\title{Integrable hydrodynamics of Toda chain: case of small systems}


	\author{Aritra Kundu}
	\affiliation{Department of Physics and Materials Science, University of Luxembourg, L-1511 Luxembourg}
		\affiliation{SISSA, via Bonomea, 265 - 34136 Trieste Italy}

	
	\date{\today}
	\begin{abstract}
		
		Passing from a microscopic discrete lattice system with many degrees of freedom to a mesoscopic continuum system described by a few coarse-grained equations is challenging. The common folklore is to take the thermodynamic limit so that the physics of the discrete lattice describes the continuum results. The analytical procedure to do so relies on defining a small length scale (typically the lattice spacing) to coarse-grain the microscopic evolution equations. Moving from the microscopic scale to the mesoscopic scale then requires careful approximations.
		In this work, we numerically test the coarsening in a Toda chain, which is an interacting integrable system, i.e. a system with a macroscopic number of conserved charges. Specifically, we study the spreading of fluctuations by computing the spatiotemporal thermal correlations with three different methods: (a) Using microscopic molecular dynamics simulation with a large number of particles; (b) solving the generalized hydrodynamics equation; (c) solving the linear Euler scale equations for each conserved quantities.
		Surprisingly, the results for the small systems (c) match the thermodynamic results
	    in (a) and (b) for macroscopic systems. This reiterates the importance and validity of integrable hydrodynamics in describing experiments in the laboratory, where we typically have microscopic systems. 
		
	\end{abstract}
	
	\maketitle
	The hydrodynamic theory is a cornerstone for understanding coarse-grained phenomena in
	many-body systems. The local evolution of slowly changing fields of conservation laws manifests the large-scale space-time dynamics of the system. This old theory has found new interest to predict transport in low-dimensional systems relevant to the current technologies. In nonintegrable systems, with few relevant conservation laws, the non-linear fluctuating hydrodynamic theory helps to understand diffusive and non-diffusive transport \cite{spohn2016}. In integrable systems with many  conservation laws, transport is historically pictured as an underlying quasiparticle undergoing deterministic scattering. Recent experiments have made novel classical and quantum dynamics in these systems \cite{kinoshita2006}. However, such a formalism can only sometimes allow for the analytical tractability of physically relevant quantities measured in the lab or computer experiments. Integrable models such as harmonic chain and hard point particle gas (or models reducible to them) have been studied within the standard kinetic theory formalism to connect this quasiparticle picture and to understand coarse-grained transport properties \cite{bulchandani2019}. The assumed simplicity of a many-body integrable system only sometimes implies the analytical tractability of physically relevant quantities. The complexity arises for interacting systems such as the XXZ chain or the Toda chain \cite{spohn2018}. The solvability in these systems are addressed using Lax pairs in classical  
	\cite{Arutyunov2019} and using the Bethe ansatz \cite{yang1969} in quantum setup but characterizing ,
    dynamical properties remain challenging using these formalisms. It is typically expected that, because of the lack of chaos in an integrable system, the transport is ballistic in nature. This conjecture is verified in the context of energy diffusion in integrable systems, for example, in Toda chain, Lieb-Linegar gas, sinh-gordan model etc. \cite{theodorakopoulos1984}. However, exceptions can be found for example, in the integrable XXZ model which is integrable but the spin transport is non-ballistic \cite{das2019}.
    
	Early attempts to describe large-scale dynamics included merging soliton theory with kinetic theory \cite{el2005,bulchandani2017}. The general picture of quasi-particles has recently been incorporated in the development of integrable hydrodynamics that address the full spatio-temporal structure of fluctuations and beyond \cite{bertini2016}.
	There is a serious effort to extend hydrodynamic theory to the quantum domain with growing interest in both experimental and theoretical perspectives \cite{castro-alvaredo2016,ruggiero2020}. 
	Curiously, experiments which contain few atoms are well described by the hydrodynamic theory initially developed in the thermodynamic limit. This simple question has no obvious answer and has deep connections to coarse-graining and  taking the hydrodynamic limit \cite{varadhan1990,spohn2012}. For example, the 1D Bose gas in the experiment  \cite{kinoshita2006} composed of $40-250$ atoms  may not seem enough for the thermodynamic limit, but the experimental results can be explained using integrable hydrodynamics.

	In this work, we theoretically explore the question: can a small system be described by hydrodynamic theory? Our analysis is based on the classical integrable hydrodynamics of the classical Toda chain as developed in \cite{spohn2021,doyon2019,nardis2018,nardis2019}. The framework can be generalized to other integrable systems, classical or quantum \cite{schemmer2019}.
	In the quantum domain, numerical simulations with time-dependent density matrix renormalization group of a thermal expansion in the XXZ model lead to a similar conclusion: hydrodynamic predictions with smooth initial conditions are accurate, even for small system sizes \cite{bulchandani2017}.

	Here, we construct the theoretical framework to study spatio-temporal correlation in the Toda chain at equilibrium with three different but related approaches. The first uses a microscopic molecular dynamics (MD) simulation similar to \cite{kundu2016}, the second is by solving the generalized hydrodynamics equation for the Toda chain  \cite{spohn2021,doyon2019} and the third by solving the linear Euler scale equations considering all the conserved quantities. The three methods have different levels of difficulty / accessibility using finite computational resources. The MD simulations can be performed for a relatively large chain (say $N=2048$ particles) for $ \sim 3000$ computational hours, while solving the integrodifferential GHD equation in the thermodynamic limit is almost instantaneous, but requires a large time to setup the framework. Solving the Euler equation is limited to a few atoms only and is of medium complexity. En route to testing the accuracy of the three methods, we find that a small system with $N=8$ particles has a spatio-temporal correlation at equilibrium almost indistinguishable from that obtained using the GHD equation in the thermodynamic limit. 
	\section{Euler scale hydrodynamics}
	
	We discuss the classical many-body interacting integrable system in a general setting, formulate the conserved quantities, and give the prescription to construct the Euler scale hydrodynamics using their respective currents.
	
	\textit{ Integrable ingredients}: Consider a system of $N$ particles with position $q_i$ and momentum $p_i$ described by the Hamiltonian $H(\{p_i,q_i\})$. The system is integrable if the equation of motion generated by the Poisson flow $\dot{A} = \{ A,H \}$ can be written in terms of the Lax equation, where the dots refer to the time derivative. The common form of the Lax equation with the Lax pairs denoted by the $N \times N$ matrices $L^{\pm} \equiv L^{\pm} (\{p_i,q_i\})$ and $M \equiv M (\{p_i,q_i\})$, is
	\beqa
	\dot{L}^\pm = [M,L^\pm] + \lambda L^\pm  .\label{eq:Lax}
	\eeqa
	Using the cyclic property of the trace, the global conserved quantities of the system are $I^k = \Tr (L^+L^-)^k$. Note that these conserved quantities are not unique and are defined to a constant prefactor. 
	These are chosen such that the equilibrium correlation matrices for energy have an order of magnitude governed by the temperature $T$ of the system. $M$ is typically a skew-symmetric square matrix, i.e. a matrix whose transpose equals its negative. We are interested in determining the local variations of these globally conserved quantities that describe the current flowing in and out of a mesoscopic unit cell in the hydrodynamic regime.
	
\begin{figure}
		\centering
		\includegraphics[scale=0.25]{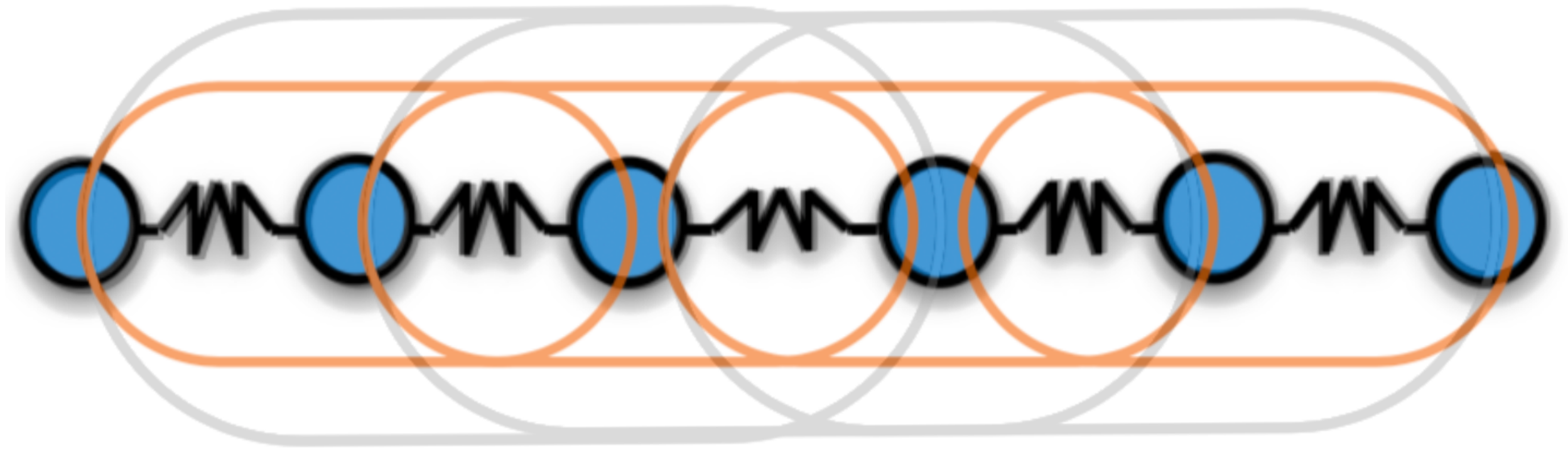}
		\caption{Schematic representation of increasing non-locality of conserved quantities of interacting integrable models with local interaction.  The orange lines show the support of two sites for a local energy, while the grey line represents a larger support for higher order conserved quantity with support of four sites.}
		\label{fig:conserved_quantity}
	\end{figure}

	\textit{Conserved quantities and currents:} The local evolution of the conserved quantity ($I^k = \sum_i I^k_i$) can be written in the continuity form by observing that they follow the same evolution under the Lax flow.  Associated with the evolution, locally conserved quantities $I^k_i$, is given as,
	\beqa
	\dot{I}^k_i = (MI^k)_{ii} -(I^kM)_{ii},
	\eeqa
	where we define the local net currents flowing in and out of the site $i$ as $ (MI^k)_{ii} = j^k_i$ and  $ (I^kM)_{ii} = j^k_{i+1}$ respectively. Typically for a $H$ with only nearest-neighbour interaction, $M$ is a two-sided off-diagonal matrix, resulting in the current flowing in and out of neighboring sites. However, for a long-range interacting system, the current flows in from all bonds. Another interesting observation is that the $I^k_i$ are increasingly non-local in real space in a short-ranged system, i.e. $I^k_i$ depend on the product of lattice variables in between $i \pm k$, pictorially shown in Fig.~\ref{fig:conserved_quantity}. This will be important to understand our results later. The eigenvalues of the matrix $L^+L^-$ is  are also conserved quantities, but, however, they are not local in real space.
	
	However, the eigenvalues of the $ L^+L^- $ matrix only capture some of the conserved quantities for translation-invariant Hamiltonian systems. There is an additional conserved quantity called the stretch variable, which is related to the inverse of the system density and plays a major role in transport. For the nearest-neighbour interaction, if the translation operator is defined as  $\{T,H\} = 0$, then $I_0 = \sum_i (T-1)q_i$  and the total momentum $I_1 = \sum_i p_i$ are the additional conserved quantity for systems with periodic boundary conditions.
	It is important to exercise some caution: Although these look like all the conserved quantities of the system, there can be others that go unnoticed, i.e. any function of the conserved quantity is also conserved. See, for example, \cite{davies2011} for the role in the non-linear Schrodinger equation also in the context of Toda lattice, which was investigated in \cite{dhar2021}.
	
	\textit{General Gibbs ensemble (GGE):} 
	The integrable system at equilibrium is described formally by GGE,
	\eqa{ P(\{p_i,q_i\}) = Z^{-1} \text{exp}[-\sum_{k=0}^N \mu_k I^k(\{p_i,q_i\})],}
	where $\mu_k$ are the dual thermodynamic variables corresponding to each conserved quantity. For example, a system at Gibbs equilibrium is described by 
	dual variable temperature ($\mu_2 = \beta$, corresponding to Hamiltonian $H$ ) and pressure ($\mu_0 = P$, corresponding to stretch or the inverse density). In the rest of the text, any average $\av{.}$ is understood
	to be taken with respect to the GGE characterized by $\beta,P$.
	
	\textit{ Euler scale (fluctuating) hydrodynamics at equilibrium:}
	 The coarse-grained large-scale space-time behaviour of the system describing the macroscopic evolution is 
	governed by local fluctuations of the globally conserved quantities. We are interested in the system prepared in a GGE characterized by canonically conjugate thermodynamic variables such as temperature, pressure, etc. In equilibrium, each conserved quantity has a well-defined average that does not change with time, i.e. $\partial_t \langle I^k_i\rangle = 0$. This contributes to the trivial shift of the conserved quantity of the system and is typically subtracted.
	The local fluctuation of the total conserved quantity $I^k$ is denoted $ u^k_i(t)= I^k_i(t) - \av{ I^k_i}$. 
	The linearized hydrodynamic Euler equations for the evolution of $u^k(x,t)$ in the continuum limit are
	
	\eqa{\partial_t{u}^k(x,t) = \partial_x j^k(x,t),}
	where we have switched from $i \to x$ in going to the continuum and $j^k(x,t)=j^k(x,t) - \av{j^k_x}$ is the local current fluctuation associated with the $k$th conserved quantity.
	Following the standard prescription of hydrodynamics to the first order in small fluctuations, we express the current in terms of the local fluctuations of the conserved fields i.e.
	$j^k(x,t) = \int A_{kl}(x,t,x',t') u^l(x',t') dx'dt'$, where the kernel $ A_{kl}(x,t,x',t') = \frac{\delta j^k(x,t)}{\delta u^l(x',t')}$ encapsulates all information of the evolution. 
	We have neglected the higher-order expansions which are not important in the ballistic scaling limit addressed in the current manuscript but become important for super-diffusive transport \cite{spohn2014,spohn2016}.
    The kernel is generally hard to compute and much simplification is obtained in the averaged current at equilibrium, where the time dependence can be dropped due to time translation invariance at equilibrium. The dependence on positions is dropped using spatial invariance and local interactions.

    We switch to vector notation, with $\mathbf{u}$ the row vector consisting of the local variation of conserved quantities and $\mathbf{A}$  a square matrix with dimensions of the number of conserved quantities. The average current over the GGE at equilibrium is \eqa{{\mathbf{j}(x,t)} = \av{\mathbf{A }} \mathbf{u}(x,t) +\partial_x \mathbf{D}(x,t) \mathbf{u}(x,t)  +\mathbf{B}(x,t) \mathbf{\zeta}(x,t). \label{eq:curr}}  Here $\av{\mathbf{A }}  = \av{\mathbf{\frac{\delta j}{\delta u}}}$ can be interpreted as the projection of the current to the conserved quantity and $\zeta$  a Gaussian white noise with $\av{\zeta(x,t)\zeta(x',t')} = \delta(x-x')\delta(t-t')$. The diffusion matrix $\mathbf{D}$ and the noise matrix $\mathbf{B}$ are added phenomenologically and cannot be easily derived from a microscopical picture. It may seem surprising that these terms are not zero here, as the chaos required for the origin of diffusion and noise is absent in an integrable system.
    Typically, the origin of these terms is argued as follows: The system  is as a fluid cell where each
	cell is a 'mesoscopic' region, i.e. a region of
	finite extent, which is large compared to the distance between particles, but small compared to the macroscopic spatial
	variation scales $L$. The dynamics inside the fluid cell can be split into two parts: the first being the net current flowing through the fluid cell. Each of these currents are projected on the basis of conserved quantities as discussed before. The 
	second part comes from the dynamics in the rapid variation in the fluid cell due to interactions with neighbour cells. This cannot be taken into account through the projection to the conserved quantities and is modelled as dissipation and noise. 
	
	This gives the Euler scale hydrodynamic equation
	\eqa{
		\dot{\mathbf{u}}(x,t) &= \partial_x \left[ \av{\mathbf{A}} \mathbf{u}(x,t) + \frac{1}{2} \partial_x \mathbf{D}(x,t) \mathbf{u}(x,t) + \mathbf{B}(x,t) \mathbf{\zeta}(x,t) \right],\label{eq:hydro}
	}
 The above equation maintains conservation laws as in the microscopic system, keeping the total fluctuation across the system constant in time.
	Although this looks formal, a more rigorous derivation using large deviation theory can be obtained along the lines of \cite{saito2021}. Further, it can be shown that (with abuse of notation, we drop the average) the matrix $\mathbf{A} = \av{\mathbf{\frac{\delta j}{\delta u}}} \equiv \la \mathbf{ju} \ra \la \mathbf{u^2} \ra^{-1}$, where we take time independent averages at GGE \footnote{A naive way to see this is to multiply the numerator and denominator by same fluctuation $u$ and taking a average}.
	The matrix $\av{ \mathbf{ju}}_{\alpha\beta}  = \av{j_\alpha u_\beta} - \av{j_\alpha}\av{ u_\beta} $, and $\av{ \mathbf{u^2}}_{\alpha\beta}  = \av{u_\alpha u_\beta} - \av{u_\alpha}\av{ u_\beta} $ does not have any time dependence and is computed at equilibrium.
	This has a nice geometrical interpretation since the change of curvature of the differential equation is interpreted as a projection of the current to the conserved quantities in the system. A more rigorous procedure will be to follow the methods presented in \cite{saito2021,spohn2021}. 
	Before going to the next section, where we discuss the ballistic correlation, a small remark is that for a non-integrable system like the Fermi-Pasta-Ulam (FPU) chain with three conserved quantities, Eq. \eqref{eq:hydro} consisting of $3
	\times 3$ matrix corresponding to the stretch, density and energy conservation. In this case, truncating the local current to linear order is not enough and one needs to keep higher-order terms. This gives rise to evolution with non-ballistic scaling and is described by the non-linear fluctuating hydrodynamic theory.
	
	\textit{Ballistic spread of correlations:
	}
	We are interested in spatiotemporal correlation functions defined as $\mathbf{C}(x,t) =\la  \mathbf{u(x,t)}\mathbf{u^T(0,0)} \ra$ which follow the  evolution as Eq.~\eqref{eq:hydro}. It is easy to check that the total correlations dictated by this evolution is constant, i.e. $\partial_t \sum_x \mathbf{C} (x,t) = 0$ as the system undergoes conservative dynamics.
	In general, the late-time evolution of the correlation functions follows a self-similar scaling from $\mathbf{C}(x,t) = \frac{1}{t^\gamma} f(\frac{x}{t^\gamma})$.
	It is expected [although not obvious] that since our system is integrable, the transport will be ballistic, that is, $\gamma = 1$. Using this in Eq.~\eqref{eq:hydro}, we see that the contribution of the diffusive term is subleading for ballistic scaling, and we can safely ignore the last term and are left with

	\eqa{\dot{\mathbf{C}}(x,t) &= \partial_x {\mathbf{A} } \mathbf{C}(x,t). \label{eq:corr}}
	
	\textit{Solving the correlation equations:}
	The evolution of the correlations is now a straightforward linear equation and
	using the Fourier transformation
	$f(x) = \frac{1}{L} \sum_{-L/2}^{L/2} f(k) e^{-\iu 2  \pi   k/L },~ x\in -L/2,-L/2+1 \dots L/2$
	and $f(k) = \sum_{-L/2}^{L/2} f(x) e^{\iu 2 \pi  x k/L },~ k\in [-L/2,-L/2+1 \dots L/2]$, 
	Eq.~\ref{eq:corr}  becomes $\partial_t \mathbf{C}(k,t) = -\frac{2 \pi \iu k}{L}  \mathbf{A} \mathbf{C}(k,t)$.
	The solution is then written as 
	\eqa{\mf{C}(k,t) = e^{-2 \iu \pi \frac{k}{L} \mf{A} t} \mf{C}(k,0) ,}
	where $\mf{C}(k,0)$ is the Fourier transform of the equal time correlations between the conserved quantities prepared in GGE. 
	The matrix $\mf{A}$ is not symmetric, however the eigenvalues are always real as expected physically from the spatiotemporal symmetries of the system. This can be undertood as follows, when the equilibrium system is flipped spatially and its currents are time-reversed, it  behaves identically to the original system. This requires that the eigenvalues of the matrix $\mf{A}$ are real. The matrix $\mf{C}$ is a symmetric matrix. 
	The inverse Fourier transform then gives the time-dependent solution
	\eqa{\mf{C}(x,t) =&  \frac{1}{L} \sum_{-L/2}^{L/2} \mf{C}(k,t) e^{-\iu 2  \pi   kx/L }.}
	Taking the continuum limit and scaling the correlation functions, this can be rewritten in scale-free form as
	\eqa{f(z)=tC\left(z=\frac{x}{t}\right) =& \frac{1}{2 \pi} \int_{-\pi}^{\pi} d{k}~ \mf{C}(0,0) e^{-\iu(\mf{A} {k} +  {k} {z})} \label{eq:scalefree1}.}
	It is convenient to go to a diagonal basis of $\mathbf{A}$, with $\mathbf{R}\mathbf{A}\mathbf{R}^{-1} = \text{diag}[\lambda]$, where $\mathbf{R}$ diagonalize the matrix. The eigenvalues of the matrix $\mathbf{A}$ are physically relevant quantities which can be interpreted as generalized velocities of ballistic propagating modes. In FPU-like non-integrable systems, with three conserved quantities,  the matrix $\mathbf{A}$ is for example the $3 \times 3$ matrix: the eigenvalues give the sound speed of the system along with a zero eigenvalue, which corresponds to the stationary heat peak of the system. In this case, the Euler approximation does not reveal the structure of the peaks itself, and one must go to higher order expansions in the current to reveal the structure of the peaks. In an integrable system with ballistic scaling, the situation is interesting, as there is only one timescale. The system of equations gives rise to $N$ nonzero velocities corresponding to the number of conserved quantities in the system.
	The eigenvalues are ordered in the sense that $\lambda_1 \le \lambda_2 \dots \le \lambda_N$ and form a linear space. The spectral gaps between the eigenvalues are smaller toward the edges and larger in the bulk. Taking these into account renormalizes the amplitude of the scale-free function at the $k^{th}$ site by $\lambda_{k+1} - \lambda_k$.
	The scale-free equation Eq.~\eqref{eq:scalefree1} can be written in terms of these velocities as,
	\eqa{
		f_{ij}(z)=& \sum_k  R_{i,k} (R^{-1}C)_{k,j}\frac{\delta({z}-\lambda_k)}{\lambda_{k+1} - \lambda_k} .\label{eq:scalefree}
	}
	The normalization can also be understood from a more physical point of view: Since the integrable dynamics of the system is conservative, normalization eliminates the  gauge degree of freedom from the eigenvectors, and we have the sum rule discussed above, $ \sum_z \mathbf{f} (z) dz= \mathbf{C}(0,0)$, where the $dz = \lambda_{k+1} - \lambda_k$ are the discrete inverse density of the ordered eigenvalues and the right-hand side is the value of correlations computed at equilibrium. The density fluctuations are studied under the ``unfolded" spectral gap of the eigenvalues of the integrable system akin to that used in random matrix theory. A discussion of this formula for the Toda lattice is shown in Sec.~\ref{sec:toda}. As we shall see, the above formula for a small system size, i.e. when $N$ is of the order of $O(1)$ reproduces the thermodynamic limit spatio-temporal correlations. Note that we have not taken any hydrodynamic or large system size limit in the above formula. This makes us wonder if the dynamics of small systems at equilibrium are sufficient to be described by hydrodynamics. 
	\begin{figure}
		\centering
		\includegraphics[scale=0.25]{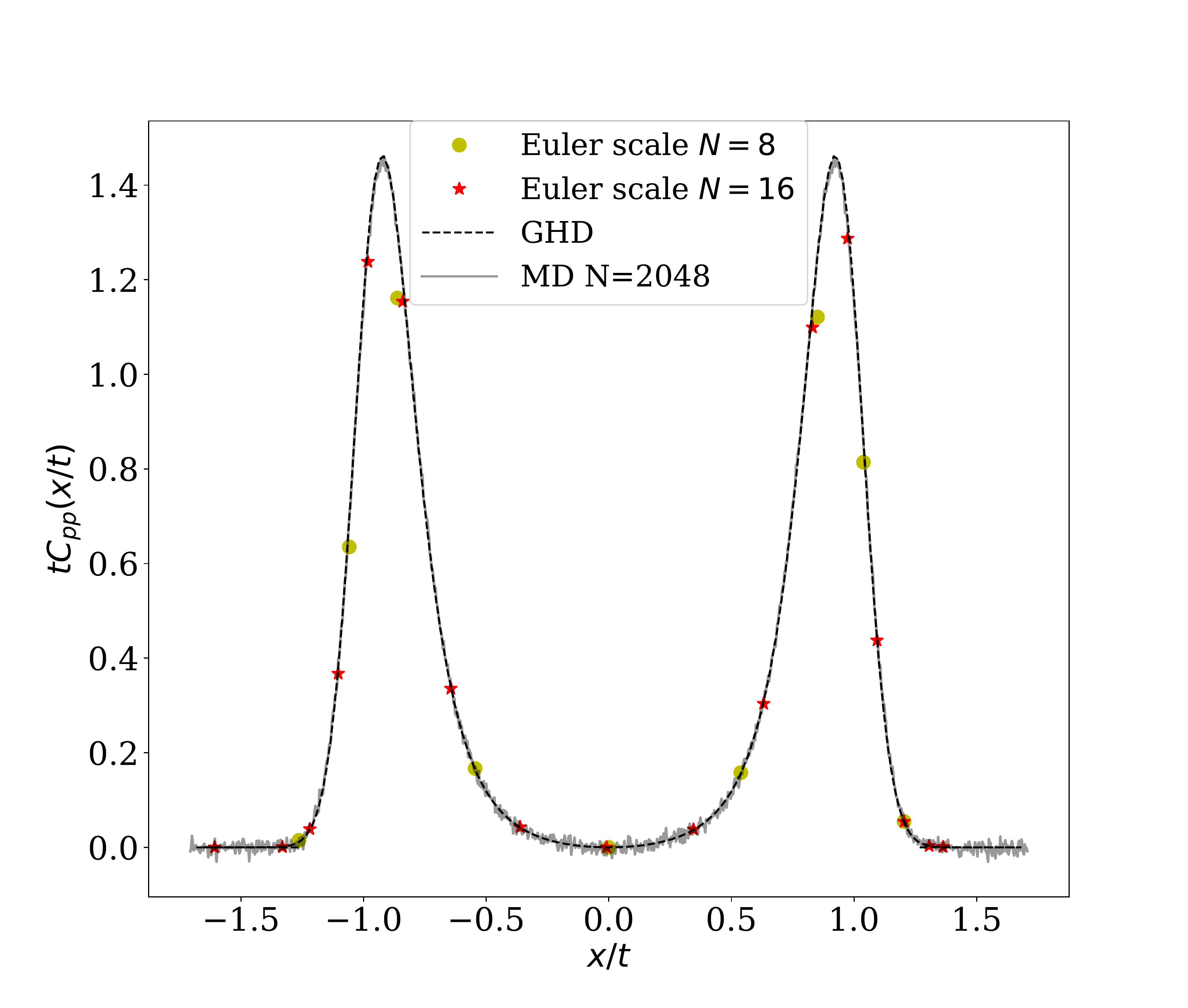}
		\caption{Comparison of scaled equilibrium correlations in a small finite sized Toda lattice with thermodynamic limit results.  The circle and the stars are data from numerically solving the Euler scale hydrodynamic equation at Gibbs equilibrium as in Eq.~\ref{eq:scalefree} for small system sizes with $N=8$ and $N=16$. The noisy grey line is microscopic molecular dynamics simulation of the macroscopic number of $N=2048$ particles with Toda interaction with an initial condition average of $10^7$ initial conditions from the Gibbs distribution with $T=1, P=1$. The black dashed line is obtained by numerically solving the thermodynamic GHD equation. We see that the results of the finite short system already have an excellent overlap with the exact results in the thermodynamic limit.}
		\label{fig:toda_correlations}
	\end{figure}

	\section{Generalized hydrodynamics: GHD}
	Generalized hydrodynamics assumes that the dynamics of an integrable system can be described by a quasi-particle which undergoes ballistic diffusion and a drift with effective velocity. At thermal equilibrium, this
	is  obtained by averaging the extensive number of conserved charges and their dynamics in a local GGE state. The natural hydrodynamic fields are the
	density of states (DOS) of the Lax matrix and the stretch, $r$, which we anticipated earlier. 
	
	For completeness, we first briefly review the basic concepts of the GHD in classical systems following the notation introduced in \cite{mendl2022,spohn2021}.
	The DOS is represented  as
	$r\rho_p(r)$, where the distribution $\rho_p$  encodes all information of the conserved quantities. Then the
	generalized hydrodynamic equations are
	
	\eqa{
		\partial_t r + \partial_x q_1 = 0, \quad \partial_t (r \rho_p )+ \partial_x \left( (v^{eff} - q_1) \rho_p \right) = 0, 
	}
	where $q_1 = r \int dw~ w \rho_p (w)$ is the average momentum of the system. The effective velocity is given by the solution of the linear integral equation of the form
	\eqa{
		v^{\text{eff}}(r) = r + (T \rho_p v^\text{eff}) (r) - (T \rho_p) v^\text{eff} (r),
	}
	with the operator $T$ being the integral of kernel ($K$) which takes into account the two-particle scattering shift. It is defined as an integral transform on a function $\psi(w)$ as,
	\eqa{ T \psi(w) = \int K(w-w') \psi(w') ~dw', }
	where the kernel $K$ depends on the microscopic model and can be computed by solving the large time asymptotic in two-particle scattering. 
	
	It is convenient to use a non-linear transformation to go to a coordinate frame, where the hydrodynamic equation becomes simplier. This is done by transformation to normal mode DOS $\rho_n$,
	\eqa{ 1-T\rho_n(r) = {{(1+T\rho_p(r)})^{-1}}.}
	
	%
	Key to this description is to introduce a dressing transformation that transforms the real valued function $\psi$ as
	\eqa{
		\psi^{dr} = (1-T \rho_n)^{-1} \psi.
	}
	In this notation, the transformation between the normal mode and the effective velocity is given as
	\eqa{ \rho_p = \rho_n [1]^{dr},~~ v^{eff}  = \frac{[w]^{dr}}{[1]^{dr}},}
	and the normal mode hydrodynamic equation is 
	\eqa{
		r \partial_t \rho_n + (v^{eff} - q_1 ) \partial_x \rho_n = 0,
	}
	It is convenient to denote the average over DOS as $\av{f}_{p} = \int f(w) \rho_p(w) dw$.
    In this notation, the conserved quantities are given as the integral identities
	$
	\av{r}_p = 1
	$ and $\av{r w^k}_p = I_k$. We are interested in the correlation function of the conserved quantities,
	
	\eqa{ C_{\alpha\beta} = [\mathbf{C}]_{\alpha\beta}  = \av{r (w^\alpha - I^\alpha[1]^{dr})(w^\beta - I^\beta[1]^{dr})}_p .\label{eq:corrnormal}}
	
	\textit{Thermal states:}
	For thermal states, the $\rho_n$ is the solution of the thermodynamic Bethe-Ansatz (TBA) like equation characterized by the chemical potential $\mu$ a function of the external pressure and temperature determined at equilibrium by condition $\int n(w) dw = P$,
	
	\eqa{ V(w) - 2 (T \rho_n)(w) + \log(\rho_n(w)) - \mu = 0, \label{eq:TBA}}
	where  $V(w) = \frac{1}{2} \beta w^2$  for the initial state in Gibbs equilibrium. Note that for a classical system there is no TBA, but this equation is obtained in Toda chain in a similar spirit using a nontrivial mapping to random matrix theory with the kernel given by $K(x) = \log(x)$, which is determined from the two-particle scattering.
	The solution to this highly non-linear root equation is often supplemented with solving the corresponding Fokker-Planck equation whose steady state solution satisfies the TBA equation above,
	\eqa{
		\partial_t \rho(w,t) = \partial_w \left( V'(w) \rho(w,t) - 2 P \int K'(w-v) \rho(v,t) \rho(w,t) dv \right) + \partial_w^2 \rho(w,t), \label{eq:FPeqn}
	}
	where $K'$ denotes the derivative of the integral kernel $K$ which is specific to the model.
	As a consistency check, the solution to Eq.~\eqref{eq:FPeqn} is inserted into the TBA Eq. \eqref{eq:TBA}, gives the chemical potential $\mu$. At equilibrium, $\mu$ turns out to be exactly the free energy when we set $w \to 0$ which acts as a cross-check for the results with the explicit formula available \cite{kundu2016,takayama1986}.
	At this point, it is convenient to introduce the quasi-energy parameterization of the normal mode obtained from the solution of the Fokker-Planck solution of TBA by $\epsilon(w) = - \log [\rho_n(w)]$. Numerically, this step is carried out by first obtaining $\epsilon$ from the Fokker-Planck equation and then plugging it into the TBA Eq. ~\eqref{eq:TBA} and finding the roots of the equation.
	The average momentum in terms of dressed variable is 
	\eqa{\av{p}_{\beta,P} = \av{r}_{\beta,P}\int \rho_n [w]^{dr} dw.}
	The correlations in Eq. \eqref{eq:corrnormal}   
	in case of momentum correlation is conjectured as [Eq.~3.17 in  \cite{spohn2021}],
	
	\eqa{ C_{pp} \approx \av{r}_{\beta,P}  \int dw  \rho_p(w)\delta(x-t{v^{eff}(w)}{\av{r}_{\beta,P}^{-1} })  ([v]^{dr})^2(w) \label{eq:GHDcorrp}.}
	Other higher-order correlations can be computed appropriately by generalising this formula.
	
	\section{Toda chain}\label{sec:toda}
	
	We now focus on applying the methods described on the 
	Toda chain \cite{toda2012} which is a model of classical particles with exponentially decaying nearest-neighbour interactions. This is an example of a classical interacting integrable particle system \cite{spohn2018} which is solvable by Lax pairs as described before. A large Toda lattice is used as a discrete model for solitonic waves while a small chain of three particles can be mapped to the famous Hennon Heiles problem.
	The history of understanding dynamics in Toda lattice is decades old: The equilibrium dynamical properties in large Toda chains were extensively studied using various methods. 
	Numerically, correlations have been explored in the literature in \cite{kundu2016} using molecular dynamics simulations. 
	In an integrable system of $N$ particles, an equivalent number of conserved quantities is enough to describe the dynamics completely. \footnote{However, some recent results challenge this assumption\cite{dhar2021}.}
	
	The Toda chain is defined as $
	H({p_i,r_i}) = \sum_i e_i = \sum_i \frac{p_i^2}{2} +  e^{-r_i}  $
	with $r_i = q_{i+1}-q_i$. The
	equations of motion (e.o.m.) are:
	\begin{eqnarray}
		\dot{r_i} &=& p_{i+1} - p_i,\nonumber\\
		\dot{p_i} &=& e^{-r_{i-1}} - e^{-r_i} .\label{eq:eom}
	\end{eqnarray}
	These e.o.m. along with periodic boundary conditions $r_{N+1} = r_{1}$ and $r_0 = r_N$ can be cast to a Lax matrix as in Eq.~\eqref{eq:Lax}, with $\lambda = 0$ and  $L=L^+ = L^-$.
	The symmetric matrix $L$ is a tri-diagonal matrix with diagonal elements $b_i =\frac{p_i}{2}$ and off-diagonal elements $a_i = \frac{1}{2}e^{-r_i/2}$.  The conserved quantities are trace of the powers of Lax matrix, denoted as $Q_n$ where $Q_1 = \sum_{i=1}^N p_i$ and $Q_2 = \sum_{i=1}^N e_i$. In \cite{shastry2010}, explicit higher-order conserved quantities are remunerated. Importantly, in real space, the conserved quantities have increasing non-local support in space(see Fig. \ref{fig:conserved_quantity}). However, this is not the only way to view the conserved quantities of the system; for example, the eigenvalues of the Lax matrix are also conserved under evolution. It has been suggested that the spacing between the eigenvalues of the Lax matrix provides a natural transition to normal mode frequencies in \cite{goldfriend2018,goldfriend2019}. These conserved quantities are an adiabatic invariant for small non-integrable perturbations \cite{grava2020}. Prior attempts have been made to understand the dynamics of the Toda chain. The inverse
	scattering method is useful in an isolated system and provides a way to address not only the solitonic properties but also the large-scale hydrodynamic properties, as discussed earlier.   Thermalization properties and the conundrum of the local observable are related to Gibbs or General Gibbs is addressed in \cite{baldovin2021}. The majority of the works address single-point functions and 
	finite temperature
	dynamical spatio-temporal properties although difficult to access has been attempted through 
	non-interacting soliton gas analogy \cite{diederich1981} and through quantum-classical correspondence with
	quantum Toda chain \cite{cuccoli1993}. Earlier attempts at
	dynamic properties of the Toda chain are reviewed in \cite{cuccoli1994}. After a gap in interest in Toda dynamics, recent development in the integrable hydrodynamic theory
	of Toda chain in \cite{spohn2016,doyon2019}, where the dynamics of single-point functions starting from domain wall initial conditions
	was addressed \cite{mendl2022}. Accurate numerical results for the dynamics of two-point functions were discussed in \cite{kundu2016} and analytical results were only available in the limiting cases when the 
	model was solvable. As a by product, this work fills the gap in theoretical description of the numerical results in spatio-temporal correlation functions described in \cite{kundu2016} from the theory developed in \cite{spohn2021,doyon2019}. 
	
	\textit{Results:}  Fig.~\eqref{fig:toda_correlations} summaries the main numerical results of computing the momentum spatio-temporal correlations at Gibbs thermal equilibrium. The system described by Eq.~\eqref{eq:eom} is solved together with the initial conditions at equilibrium described by temperature ($\beta^{-1} = T$) and pressure ($P$) in a Gibbs ensemble with $ Z = \prod_i e^{-\beta (e_i + P r_i)}$. The noisy grey line is a microscopic numerical simulation of the macroscopic number of $N= 2048$ particles with Toda interaction with an average of $10^7$ initial conditions sampled from the Gibbs distribution with $T=1, P=1$. The black dashed line is obtained by numerically solving the GHD equation in Eq.~\eqref{eq:GHDcorrp} with the input from the TBA equation for computing the effective velocity of the quasi-particle and analytically computed thermal expectation values. Exact numerical details of the relevant computation is given in \cite{mendl2022}, and a code is available at \cite{mendlcode}. The circle and the stars are data from numerically solving the Euler scale hydrodynamic equation at Gibbs equilibrium as in Eq.~\eqref{eq: scale-free} for small system sizes with $N=8$ and $N=16$. As anticipated earlier in the introduction, surprisingly, even for the small system sizes, the spatiotemporal correlations have a good agreement with correlations in a large system sizes. This is indeed a surprising observation, and it make us wonder about the question of taking a thermodynamic limit, which serves as the motivation of the problem. Note that the positions for each of the dots corresponding to solving the Euler scale equation can be interpreted as ``generalized" velocities of the system in the ballistic scaling limit. When the dimensions of the Lax matrix are of order $N$, there are $N+1$ velocities, which overlaps with the thermodynamic solution of the GHD. It is interesting to note that these generalized velocities are neither symmetrical nor are they equally spaced. The peak of the curves are not captured well with this method, however, the possibility of the effects of taking of even-odd number of particles cannot be ruled out. The relative density of state of the distribution of the Lax eigenvalues are shown in Fig.~\eqref{fig:doseval}, which shows that the eigenvalues are closer to the edges and more spread out in the bulk. 
	
\begin{figure}[H]
		\centering
		\includegraphics[width=0.45\linewidth]{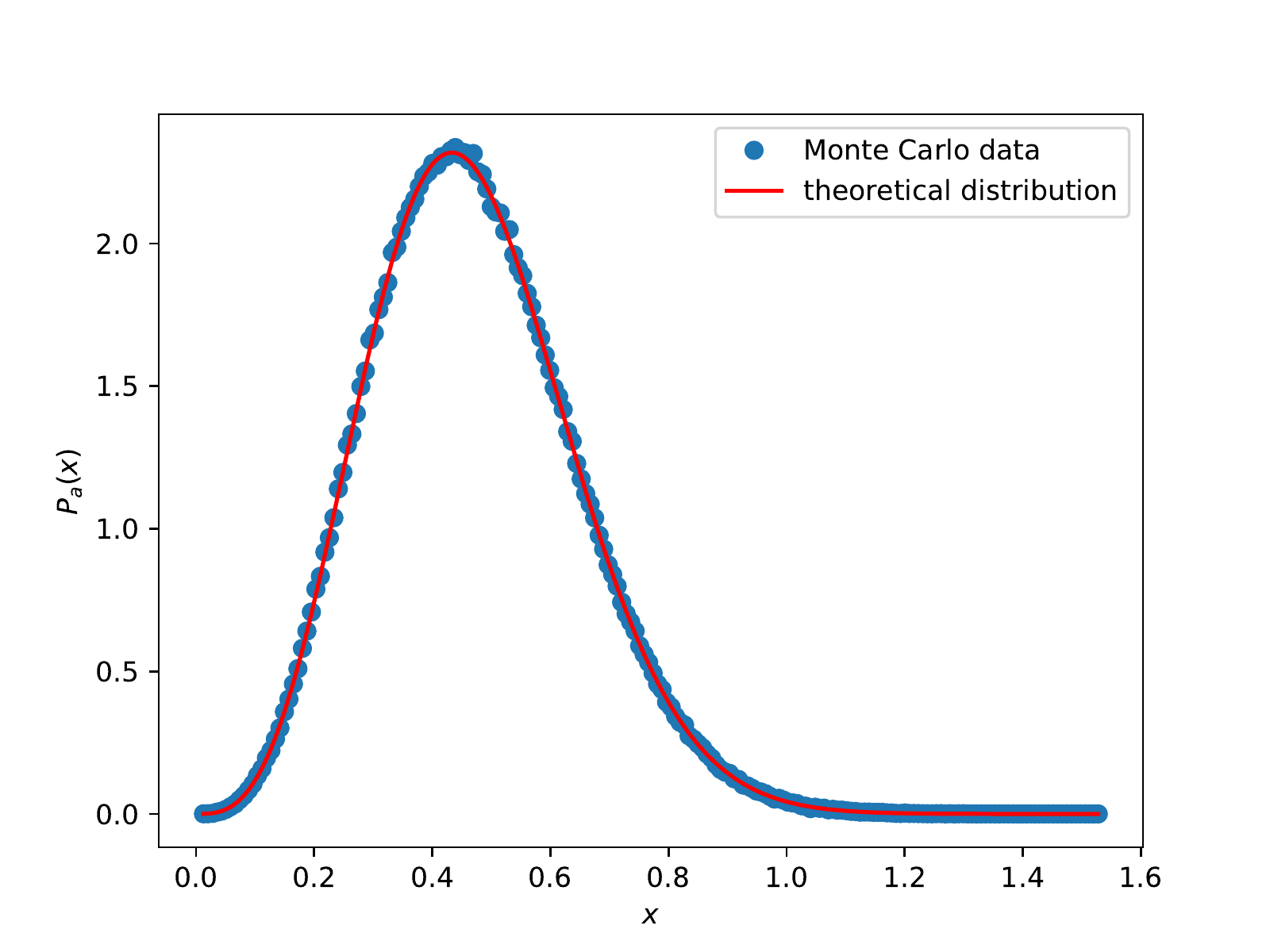}
		\caption{Monte-carlo sampling of the distribution of the off-diagonal elements of Lax matrix as given by equation Eq.~\ref{eq:offdiag}. We use this to compute the Lax matrix's eigenvalues and the correlation functions.}
		\label{fig:adist}
	\end{figure}

 	{\textit{Numerical method for direct Euler scale dynamics:}}
	The numerical method used to simulate the hydrodynamics of the Euler scale is by Monte Carlo sampling of the Lax matrix. The trick here is to do the Monte Carlo sampling from the Lax matrix itself corresponding to the Gibbs ensemble of the system at a specific temperature and pressure. For the Toda system, it can be shown that in Gibbs distribution: i.e. when the system is prepared in $P(\{p_i,r_i\}) = Z^{-1}e^{-\beta (H(\{p_i,r_i\}) + P r_i)}$ ($Z$ is the normalization), the elements $a_i  $ follow the chi-square distribution (Fig.~\ref{fig:adist} for numerical sampling),
$P_a(x) = \frac{2 (4 \beta)^{\beta P}}{\Gamma(\beta P)}x^{2\beta P -1}e^{-4 \beta x^2} ,\label{eq:offdiag}$ and the momentum follow the Gaussian distribution. The eigenspectrum of the Lax matrix is used to compute the Euler-scale hydrodynamics along with the equilibrium correlations of conserved quantities. In our simulations, we used a  $10^9$ sample of the Lax matrix and checked the convergence of the eigenvalues of the matrix $\mathbf{A}$. The equilibrium correlations of the conserved quantities are more difficult objects to deal with analytically. As we mentioned before, the higher-order conserved quantities have increasing non-local support in real space, and the convergence of equilibrium correlations becomes difficult. However, here we are interested in the system at Gibbs equilibrium, where the relevant conserved quantities that enter the distribution are energy and stretch, which are both extremely local in space. This makes the connected correlation of the higher-order conserved quantities negligible compared to the correlations of energy and the stretch variables. Interestingly, truncating the Lax matrix to include only the first three conserved quantities which appear in the Gibbs ensemble: the energy, momentum, and the stretch reproduce to $5\%$ error on the motion of the peaks of the dynamical fluctuations \cite{kundu2016} with the non-zero eigenvalues $\lambda= \pm 0.8833$ and we conclude that the correlations of the higher non-local conserved quantities are important to capture the dynamical correlations of the more local conserved quantities.
	
	{\textit{Comparison and discussion and limitations of results}}   
	
	We reiterate that these results are strictly for correlations in the ballistic scaling limit. There are some intuitive results on the nature of diffusive corrections on the ballistic scale, which we will not discuss here \cite{doyon2019}. Although the method seems general, it is by no means more efficient numerically than solving the GHD equations directly. The GHD framework provides a general analytic framework for working with set generic initial conditions. On the other hand,  the Euler scale hydrodynamics is a more brute-force approach to the problem using Monte Carlo sampling of the initial data and is limited to small system sizes.
	
	\section{Conclusions}
	We have discussed the adequacy of using integrable hydrodynamics to describe small classical (and quantum) systems. We show that although the hydrodynamic theory is developed for a large number of particles in thermodynamic limit, practically, using an independent direct method discussed here, a relatively small system size in a classical interacting integrable system is already quite accurate in describing the results in thermodynamic limit. We have tested this for a short-ranged interacting system; however, questions related to the finite-size hydrodynamics in a long-ranged system (for example in Colegaro Sutherland model) remains to be explored.
	
	\textit{Acknowledgements:} I thank Herbert Spohn,  Abhishek Dhar, Manas Kulkarni for discussions. I thank Christian Mendl for his correspondence and GHD code. The conference SPLDS2022 in Ponte-e-Mousson on the occasion of the birthday of Malte Henkel in 2022 served as a transition point for writing this draft. After completion of the article, I was informed that a similar study involving MD of is being addressed in \cite{grava22}, due to a conflict of interest, we decided to publish the articles separately. I thank Aurelia Chenu for reading and improving the quality of the paper. This work was partially funded by the Luxembourg National Research Fund (FNR, Attract grant 15382998).

\newpage
	\section{Appendix}\label{sec:app}
\begin{figure}[H]
		\centering
		\includegraphics[width=0.45\linewidth]{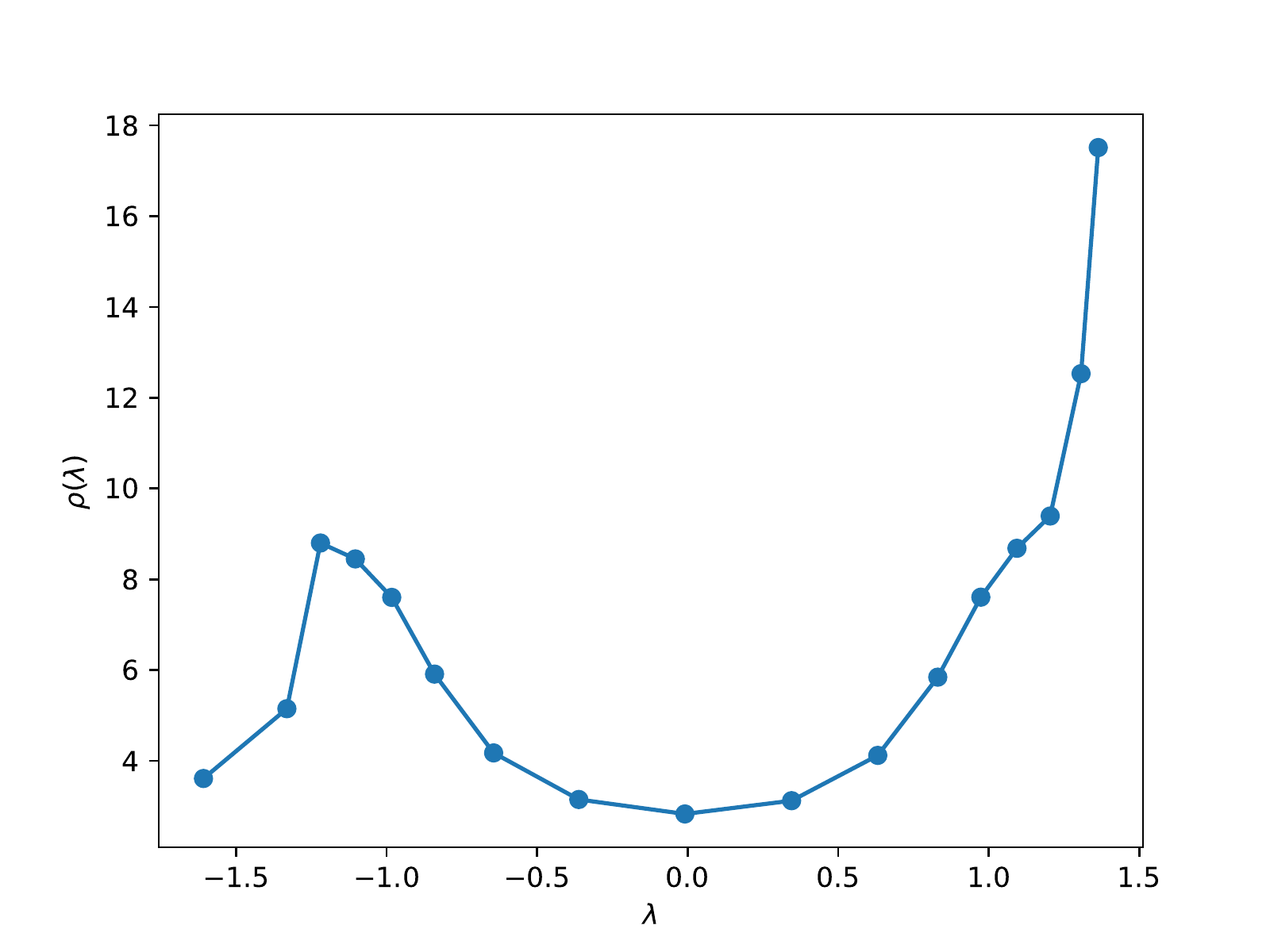}
		\caption{The denisty of states $1/(\lambda_{i+1} - \lambda_i)$ of the distribution of sorted eigenvalues of the  Lax matrix for the Toda lattice in Gibbs equilibrium at temperature $T = 1$ and pressure $P=1$. The sampling of the Lax matrix at equilibrium is achieved through transformation of the probability distribution. The diagonal elements of the Lax matrix are Gaussian to sample while the off-diagonal elements  is shown in Fig.~\ref{fig:adist}}
		\label{fig:doseval}.
	\end{figure}

	\textit{Density of states:} The relative density of states of the finite-dimensional Lax matrix for Toda used to calculate the correlation function is shown in Fig.~\ref{fig:doseval}.

\textit{ Monte-Carlo simulation of the Lax matrix:\\
}	
 \textit{Numerical values in matrix $\mathbf{A}$:}

The matrix  $\mathbf{A}$ for $N=16$ truncated to $4\times 4$ matrix:
\[
\mathbf{A}=\begin{bmatrix}
  -0.0002 & -0.9993 & 0.0017 & -0.007&\dots\\\
  -0.3357 & 0.0021 & 0.4819 & -0.0114\\
  0.0002 & -0.2333 & -0 & 1.2495\\
  -0.234 & 0.0038 & -0.0556 & -0.0313\\
  \vdots&&&&\ddots
\end{bmatrix} 
\]

\end{document}